\documentclass{IEEEcsmag}

\usepackage[colorlinks,urlcolor=blue,linkcolor=blue,citecolor=blue]{hyperref}
\expandafter\def\expandafter\UrlBreaks\expandafter{\UrlBreaks\do\/\do\*\do\-\do\~\do\'\do\"\do\-}
\usepackage{upmath,color}
\usepackage{epigraph} 

\jvol{XX}
\jnum{XX}
\paper{8}
\jmonth{Month}
\jname{Publication Name}
\jtitle{Publication Title}
\pubyear{2023}

\begin{document}

\sptitle{Theme Article: Special Issue on Security and Privacy for the Metaverse}

\title{Truth in Motion: The Unprecedented Risks and Opportunities of Extended Reality Motion Data}

\author{Vivek Nair}
\affil{University of California, Berkeley, CA, 94720, USA}

\author{{Louis Rosenberg}}
\affil{Unanimous AI, Pismo Beach, CA, 93448, USA}

\author{James F. O'Brien}
\affil{University of California, Berkeley, CA, 94720, USA}

\author{Dawn Song}
\affil{University of California, Berkeley, CA, 94720, USA}

\markboth{Special Issue on Security and Privacy for the Metaverse}{Special Issue on Security and Privacy for the Metaverse}

\begin{abstract}
Motion tracking “telemetry” data lies at the core of nearly all modern extended reality (XR) and metaverse experiences. While generally presumed innocuous, recent studies have demonstrated that motion data actually has the potential to profile and deanonymize XR users, posing a significant threat to security and privacy in the metaverse.
\vspace{1em}
\end{abstract}

\maketitle

\chapteri{W}hile virtual reality (VR) has been around in some form since well before the modern internet, the recent introduction of affordable standalone VR devices, such as the Meta Quest 2, has marked a turning point in the accessibility of VR to average consumers. In 2022 alone, more than 10 million VR headsets were sold, demonstrating that the technology has begun to reach mass-market adoption.
While the use of augmented reality (AR) devices, such as the Microsoft HoloLens, Meta Quest Pro, and upcoming Apple Vision Pro,  currently lags behind VR, AR technology is now being used in a growing number of industries and professional applications.

VR and AR, collectively known as ``extended reality'' (XR), are envisioned by their proponents as a step towards the eventual creation of a massively connected ``metaverse'': an immersive virtual world where users meet to work, learn, and socialize. Indeed, future iterations of these devices, particularly those that support AR, are well-positioned to become a major medium of human-computer interaction in the near future.

While modern XR devices contain a wide variety of sensors, at the core of nearly all XR experiences is a stream of motion capture ``telemetry'' data that records the position and orientation of tracked locations on the user's body in 3D space. Metaverse platforms, by their very nature, turn every movement of a user into a stream of data broadcast to other users anywhere in the world in order to facilitate real-time interaction.

Today's XR platforms and experiences have been built under the assumption that this telemetry data is relatively innocuous: useful for rendering an avatar representing one user on another's device, but not much more. However, a growing body of academic research challenges this notion.

In this article, we'll discuss recent studies that paint a very different picture of motion data. What appears at first to be random variations in movement may perhaps be more akin to a DNA sequence, revealing the identity, biometrics, demographics, and even health information of XR users to anyone else in the same virtual world.

The privacy consequences of XR motion data are more striking still in light of how these devices are actually used in practice. While proponents emphasize brand-friendly work meetings and social gatherings, XR usage today often includes rowdy gaming sessions or adult experiences. The ability to link user identities across applications, and perhaps even to their real-world identity, could entail severe consequences for ordinary XR users and tarnish the reputation of metaverse technologies as a whole.

The news is not entirely negative. We are still in the early days of XR adoption, and have the opportunity to learn from decades of security and privacy research on the conventional internet. In addition to describing the security and privacy challenges presented by XR motion data, we propose several clear approaches for counteracting these threats. If researchers act quickly to design and test privacy-preserving mechanisms for the metaverse, security and privacy can be at the foundation of future metaverse systems.

\eject

\epigraph{``Cassius: 'Tis Cinna; I do know him by his gait; He is a friend.''}{William Shakespeare in \textit{Julius Caesar}, 1627}

\section{TRUTH IN MOTION}

Most people have an intuitive understanding that the way we move around in our daily lives is as much an expression of our individuality as is the way we speak.
Because movement patterns are a product of each individual's unique physiology, muscle memory, and even personality, we all learn, without really trying, to recognize people based on their motion, and to make subconscious assumptions about people based on the way they move.
Actors in film and television are well aware of this, and are often instructed to adopt specific mannerisms in their movements to convey subtle cues about the character they wish to portray.

The phenomenon of persons being characterizable by their motion patterns first became the subject of rigorous academic interest in the 1970s, with a series of studies demonstrating the extent to which individuals unknowingly reveal identifying information about themselves via their movements. Most famously, in a 1977 study of six participants, Cutting and Kozlowski demonstrated that individuals can identify their friends just by viewing the motion-tracked objects affixed to the body.$^1$ At a time well before the advent of modern computer graphics, the authors creatively resorted to taping highly reflective objects to a number of points on the participants' bodies. The scientists then streamed a camera feed of the subjects through a television monitor, and increased the contrast until the participants' silhouettes disappeared and only the individual points of light could be seen, as shown in Figure 1.

\begin{figure}[h]
\centerline{\includegraphics[width=17.5pc]{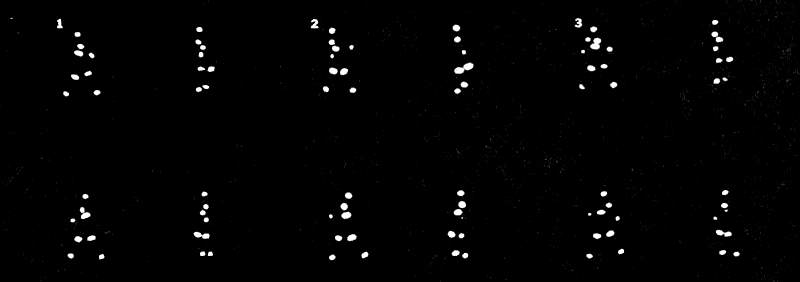}}
\caption{Three subjects are shown walking around a laboratory with point-light markers affixed to their bodies. (Adapted from `Recognizing friends by their walk: Gait perception without familiarity cues,'$^1$ with permission.)}\vspace*{-5pt}
\end{figure}

After recording the motions of six participants, their friends were asked to come into the lab and identify the name of each subject based only on the movement of the points of light, which they were able to do with 38\% accuracy (p < .005). In a later study, the same recordings were shown to a new set of participants, who were able to infer the gender of the original subjects with 79\% accuracy (p < .05).$^2$
More recently, researchers have also shown that the motion of children can be differentiated from that of adults with 66\% accuracy.$^3$

These results tell us something fundamental about human motion: it is a \textit{biometric} that belongs in the same category as blood type or an iris scan. While we have known this for some time, it is becoming particularly relevant today as we potentially enter a new era of extended reality proliferation.

\section{MOVING ABOUT THE METAVERSE}

To those who are familiar with XR, the motion data illustrated in Figure 1 may seem quite familiar.
Fundamentally, an XR device uses an array of sensors to generate a stream of motion data from its user.
As in the Cutting and Kozlowski study, XR devices function by tracking the location of individual parts of the body in 3D space. At a minimum, the location and orientation of the user's head and hands are tracked, though full-body tracking is becoming increasingly common.

In a typical consumer-grade XR system, the points of interest on the user's body are measured by the XR hardware between 60 and 144 times per second. This data is then passed to the software application running on the device, which uses it to render stimuli for the user, thereby creating an immersive experience.
In the case of multi-user or ``metaverse'' applications, the motion data is also streamed from the device to a remote game server, which in turn may forward it to other users around the world so that a virtual ``avatar'' of the first user can be rendered on their devices.

Despite changing hands several times, the information within this stream is fundamentally unchanged: individual points, representing specific body parts of the XR user, moving around in 3D space.
In other words, each of the involved entities (the hardware, the application, the server, and the other users) are receiving the same data that we have known for decades can be used to identify and profile individuals.

We are not the first to make this observation. Researchers have for some time been studying the ability to uniquely identify users based on their motions in XR. However, it is only with the recent widespread adoption of XR that sufficiently large datasets have become available to truly understand the true scale and implications of this threat.

\section{MOTION AS IDENTITY}
In 2020, a team of scientists at Stanford University's Virtual Human Interaction Lab performed an experiment to investigate whether ordinary people can be identified in VR based on their movement patterns. The researchers set up an interactive VR exhibit at The Tech Interactive, a science and technology museum in San Jose, California. Visitors to the exhibit were asked for permission to have their motion data recorded while they interacted with the VR devices being displayed.

Later, the researchers anonymized a portion of the data from each visitor to see if they could re-identify them based on their motions. The results, published in \textit{Nature Scientific Reports}, show that 95\% of users were correctly re-identified by simple machine learning models trained on less than five minutes of tracking data per person.$^4$

This is particularly noteworthy in light of the fact that the users weren't doing anything particularly identifiable; in fact, participants were just asked to passively observe $360^{\circ}$ videos while their movements were recorded. Still, in doing so, most users subconsciously revealed enough information about themselves to consistently stand out from the other 510 participants.

While this study was the first to genuinely establish the possibility of telemetry-based identification in VR, it does not tell the full story about the extent of the resulting privacy threat. For instance, identification of 511 users does not preclude the possibility that basic static measurements like height and wingspan were enough to tell each of the users apart. Further, it does not necessarily prove that users can be linked from one usage session to the next. This motivated the authors of this article to scale up the prior efforts to a size more representative of future metaverse environments.

In early 2023, we performed a similar study with data from over 55,000 users. Using over 2.5 million motion capture recordings from ``Beat Saber,'' a popular VR rhythm game, we analyzed the possibility of training machine learning models based on the gameplay recordings of each user, then identifying those users from their motions on different in-game ``maps'' and on a completely different day. Our results, to appear in \textit{USENIX Security '23}, demonstrate that users can be uniquely identified with 94.33\% accuracy from 100 seconds of data, and with 73.20\% accuracy from just 10 seconds of motion data.$^5$ In other words, by observing the movements of an anonymous VR user, we can usually determine exactly which of the 55,000 known users they are within 10 seconds, and can almost always do so within 100 seconds.

\eject

Our research in this area indicates that movement patterns, as measured by VR devices, are a much stronger biometric signal than previously imagined. It's only with the recent explosion in popularity of VR gaming that a study of this scale has become possible.
As larger datasets emerge, we may soon find that motion data can in fact identify users at an even greater rate, perhaps 1 in 100,000 or more.

\section{MOTION AS A FINGERPRINT}

To contextualize the strength of VR motion data as an identifying signal, it is helpful to compare the biometric uniqueness of motion to more traditional biometrics like iris, fingerprint, or facial scans, as shown in Figure 2.

\begin{figure}[h]
    \centerline{\includegraphics[width=18.5pc]{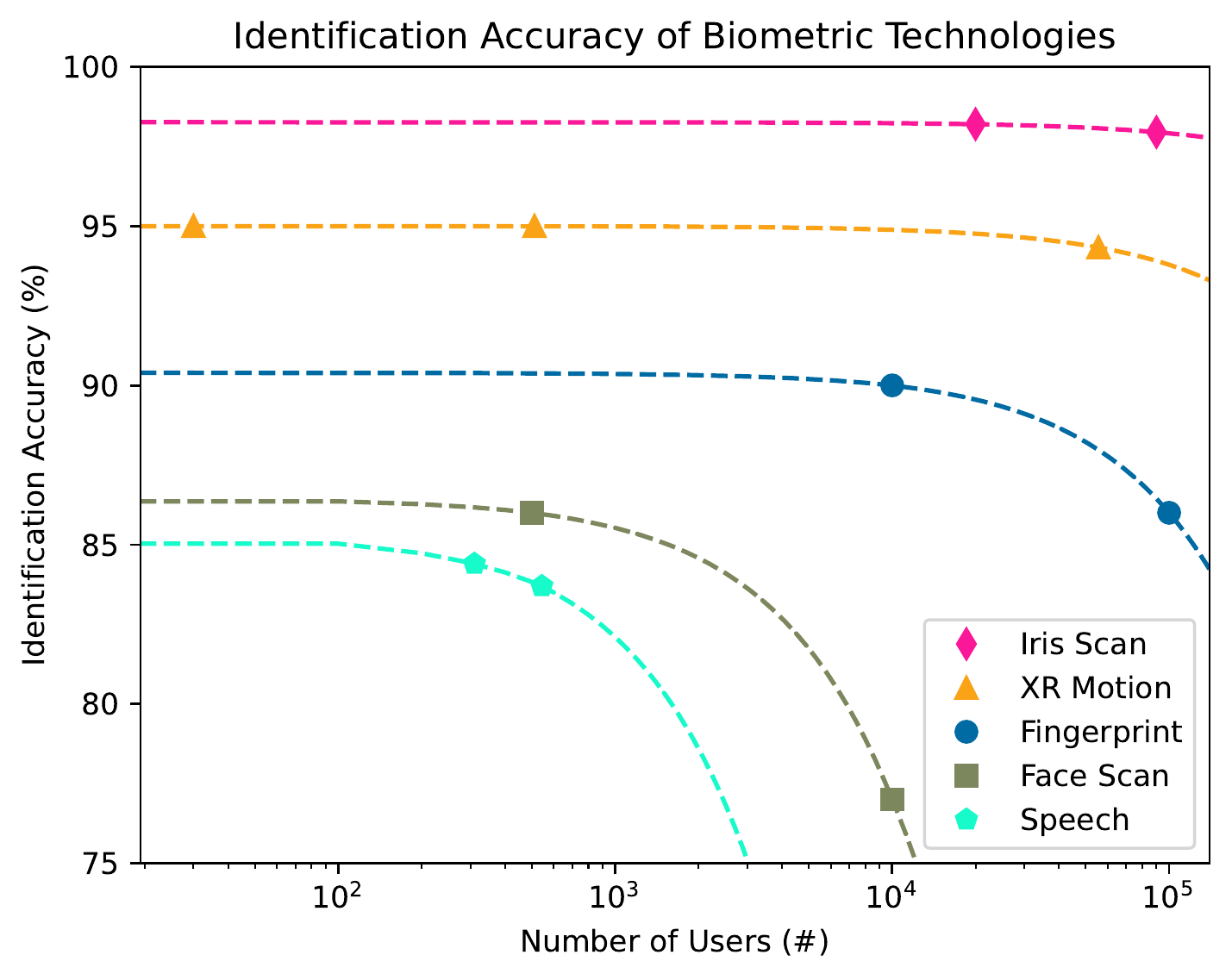}}
    \caption{Graph of user count vs. identification accuracy for various biometric technologies; log scale. (Original.)}\vspace*{-5pt}
\end{figure}

To date, the most comprehensive analysis of biometric identification is a 2003 study from the National Institutes of Standards and Technology (NIST), which analyzed dozens of commercially-available biometric sensors using real data from over 100,000 users.$^6$ The results indicate that high-end fingerprint sensors could, at the time, identify users within a population of 10,000 with 90\% accuracy. The best-performing facial recognition systems could only identify 1 in 500 users with the same accuracy. Voice recognition was shown to be even worse, with no system achieving greater than 85\% accuracy regardless of the population size.$^7$

We already know that XR motion data can be used to identify at least 55,000 users, and likely more, with over 90\% accuracy. In fact, of the technologies evaluated by NIST, only iris scans out-performed motion, achieving an identification rate better than 1:150,000.$^8$

\eject

Of course, biometric technology has greatly improved since 2003, but an equally comprehensive analysis has not since been performed. Nevertheless, the comparison remains informative; we are now in the early days of motion-based identification, and should expect to see similar improvements in motion biometrics over time. Overall, the ability to identify users via their motion is at least comparable to other biometrics at a similar stage in their developmental history.

Still, there is at least one critical difference between fingerprints and the motion data captured in XR. Sharing fingerprints, and other similar biometrics, is not strictly required to browse the web, but motion data is a fundamental part of how XR devices work, and must be shared in real time with a variety of parties to enable metaverse experiences. The equivalent would be if logging into a social media website entailed sending a scan of your fingerprints not only to the platform but also to every other user you interact with.

\vspace{-1em}

\section{A MOVING THREAT}
Consider a public figure who regularly uses a VR system with their corporate credentials to hold meetings and do professional work. In the evening, they log on with a different account to play multiplayer VR games (where they might not behave in the most professional way), and later in the evening, they use a third account for adult VR experiences.
Most people in this situation would reasonably prefer that the service providers not be able to tie these accounts together. As it stands, the user's unique motion patterns would allow any observer (or group of colluding observers) to quickly link all of these accounts to together.

On the web, ``browser fingerprinting,'' which uses subtle differences between browser configurations to link people across web services, is a highly analogous attack that is generally regarded as a significant privacy concern. However, while one can replace their browser, they cannot easily change the physiology and muscle memory that dictates their movements.

In fact, motion patterns are so intrinsically tied to our physical selves that they may soon be able to follow us out of the metaverse and into the real world. Machine learning models designed to extract 3D motion data from monocular video feeds are rapidly improving. We can reasonably extrapolate that it will eventually be possible to match a person's VR movements to surveillance video, and unlike your face, which can be covered with a mask, no reasonable countermeasure can obscure all of your movements from public view.

On the flip side, the relatively consistent nature of identifiable motion patterns could provide an unparalleled opportunity for passive authentication in future metaverse applications. XR users could benefit from the convenience of having their motion data also be used to verify their identity rather than needing to authenticate explicitly.
Unfortunately, the laissez-faire nature with which VR motion data is currently broadcasted and uploaded to the internet undermines its future use in authentication. The equivalent would be using fingerprint login on your accounts if pictures of your fingerprints were already uploaded to the internet. In a sense, today's VR users are paying a heavy early adoption penalty by sharing their motion data with the world before comprehensive defenses are in place.

Finally, like a fingerprint, one may be inclined to believe that motion identification is the virtue of random but ultimately meaningless variations. In reality, our movement style is crafted over time as the result of our background and experiences, and can later be ``decoded'' to not only identify us, but also to infer a variety of attributes that we may prefer remain private. These risks also extend to children, who will increasingly use XR devices in the coming years, not only for gaming but also for school and other educational contexts.


\vspace{-1em}

\section{MOTION AS DNA}
Thus far, we have explored the analogy of motion to a fingerprint that follows users throughout the metaverse, allowing them to be tracked across devices and applications. This analogy is true, but incomplete. Recall, for example, that point-light motion data has long been known to reveal not only the identity of participants, but also their age and gender. Perhaps a more appropriate analogy is DNA, which is not only unique to an individual, but also encodes information about their personal characteristics.

In a second study of the same Beat Saber users, we surveyed 1,006 players to ask them a variety of questions about their background, biometrics, demographics, health information, behavioral patterns, and technical device specifications. Later, we trained a series of machine learning models to see which, if any, of these responses could be accurately inferred just by examining the motion patterns of these users.$^{9}$

The results go far beyond inferring the expected anthropometrics like height and wingspan, or even demographics like age and gender. We found that even behavioral attributes, such as substance use, could be inferred from the telemetry data with statistically significant accuracy. Everything from the country that a user is from to the clothes that they are wearing can be determined using features derived from their motions alone. Perhaps most strikingly, the presence of mental and physical disabilities could clearly be discerned from the motion data. All of this -- more than 40 attributes in total -- from recordings of users playing an otherwise innocuous VR rhythm game.

With open access to device APIs, XR developers are not limited to creating such legitimate applications. Malicious developers can go further, creating games and applications that are deliberately designed to covertly reveal user data that would otherwise be hard to observe. In a third study, we recruited 50 participants to play an innocent-looking VR game called ``MetaData,'' shown in Figure 3. The game appears to be a harmless ``escape room'' experience, where players complete a series of puzzles and challenges to progress through the game. In reality, we had carefully constructed each puzzle to covertly reveal more information about the players based on their interactions with the virtual world. If ``Beat Saber'' is like a typical website, passively recording interactions in an otherwise normal application, then ``MetaData'' prototypes a concept more akin to the online quizzes deployed by Cambridge Analytica and others to actively harvest user data while being disguised as a harmless activity.

\begin{figure}[h]
    \centerline{\includegraphics[width=18.5pc]{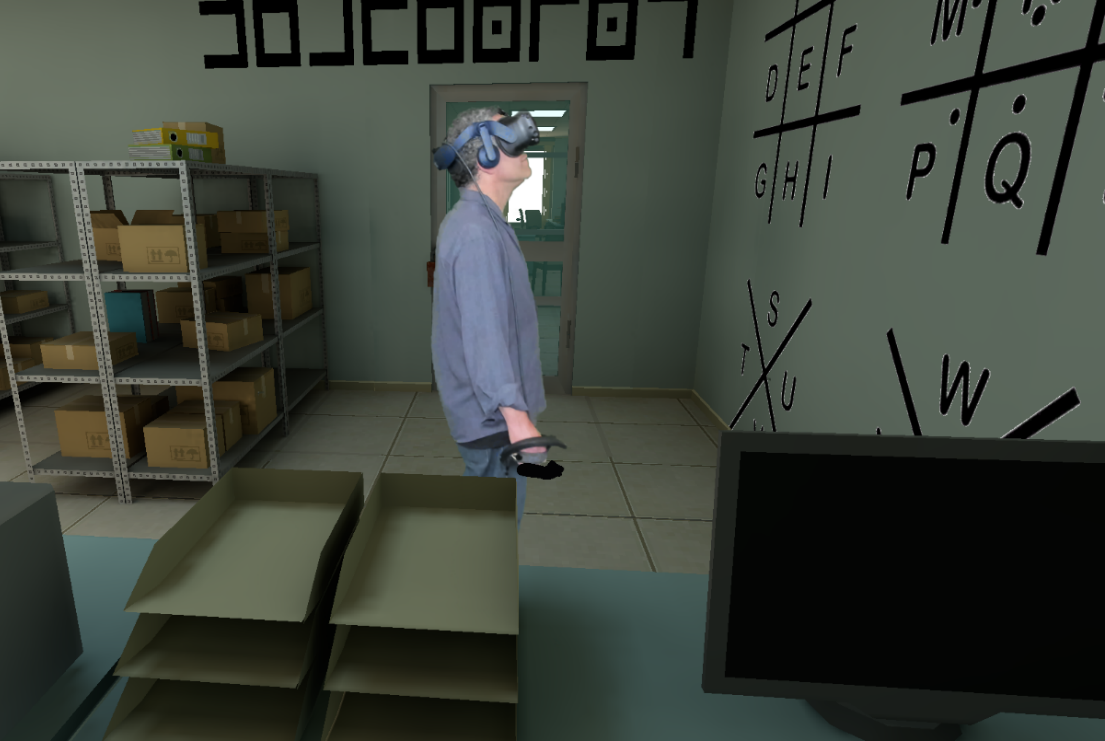}}
    \caption{Mixed reality photo of Louis Rosenberg playing ``MetaData,'' an adversarial VR escape room game. (Original.)}\vspace*{-5pt}
\end{figure}

Our results, to appear in the \textit{Privacy Enhancing Technologies Symposium}, show that over 25 personal data attributes, from anthropometrics like height and wingspan to demographics like age and gender, can be inferred from these users within just a few minutes of gameplay.$^{10}$ When combined with the identification methods discussed earlier, they can be used to aggregate user profiles from data across many applications.

These findings highlight that the privacy risk of XR devices stems not only from their sensors, such as accelerometers and gyroscopes, but also from the immersive nature of their displays, which can be used to totally control a user's virtual environment to influence the information they reveal.

\section{A FAST-MOVING FIELD}
All of the attacks we have described thus far utilize just three tracked locations: one on the user's head, and one on each hand. While that's already enough to identify and profile a large number of users, future XR headsets will likely feature full-body tracking systems, in which at least six to ten body parts are tracked.

Any risk to privacy in XR is further exacerbated by other modalities found on many devices. For example, Apple's new ``Vision Pro'' device is said to feature a LIDAR array, eye tracking, microphones, and no less than 14 cameras, in addition to full-body tracking.

As seen in Apple's upcoming device, the industry is rapidly transitioning from traditional VR headsets that are used in fully simulated worlds to lighter-weight mixed and augmented reality devices that enable immersive content to be integrated into a user’s view of their physical surroundings. The upcoming headset from Apple uses “passthrough cameras” to augment the real world with virtual content and is intended for regular use within a user’s home or office. This means users will be able to perform many of their common daily activities while wearing these mixed reality headsets, from sitting on their living room couch and opening their refrigerator to grabbing coffee mugs off the shelf or climbing into bed.  

Considering that Beat Saber data, which demands a relatively narrow range of human motions, can be used to distinguish 1 user among 55,000, we can reasonably predict that as XR devices are integrated into common daily activities, far more motion patterns will be captured, which could be used to identify and profile users with even greater precision. Consider, for example, the ubiquitous task of grabbing a doorknob and opening a door. Each of us has performed this motion so many times that it’s likely to be deeply ingrained in our muscle memory and likely at least as uniquely repeatable as the sword strikes in Beat Saber. 

Major tech companies are already developing AR and MR eyewear that they hope will be so lightweight and fashionable that users will be comfortable wearing them outside the home or office as they go about their normal daily routines: walking down city streets, shopping in retail establishments, and visiting restaurants and bars. Google, Samsung, and Qualcomm have publicly announced a partnership to develop XR devices built on the Android operating system with the goal of enabling similar usage patterns as mobile phones. In fact, many experts believe that lightweight XR eyewear will largely replace the handheld mobile phone market over the next five to ten years.  

\eject

These mobile XR devices are likely to also include an array of sensors that our interactions with the physical world. Consider, for example, grabbing products off of store shelves. Like turning a doorknob, reaching for a product is likely ingrained in our muscle memory and uniquely identifiable. However, this tracking this action is particularly interesting because mobile XR devices are capable of displaying promotional content to users based on where they are, what they’re looking at, and even what they reach for.$^{11}$ So, when picking up a can of soup, a mobile XR device could deploy targeted promotional content that links to data about that user’s personal preferences and shopping habits. Because we know XR users can be uniquely identified via their motions, it may be difficult for platform providers to maintain the privacy of users who do not wish to be individually targeted by real-time marketing materials. 

\begin{figure*}
    \centerline{\includegraphics[width=\linewidth]{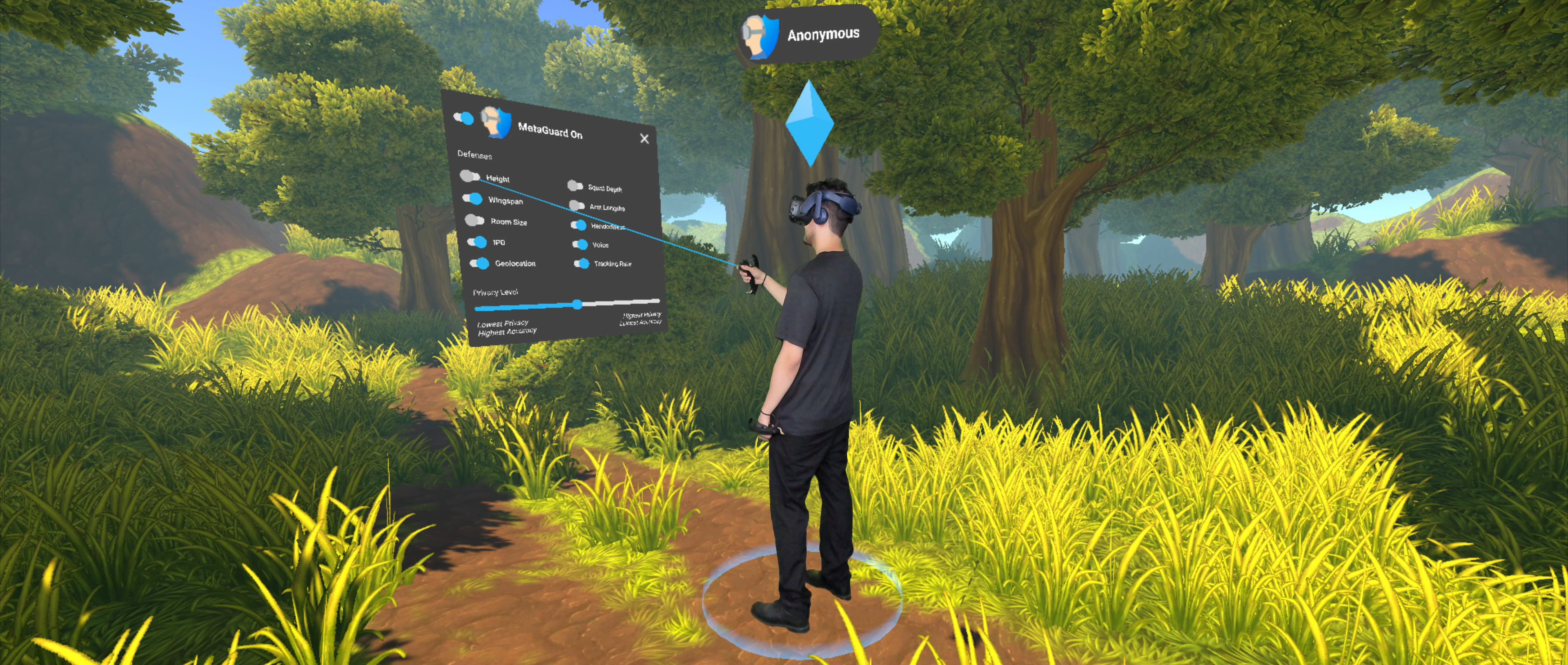}}
    \caption{Mixed reality photo of Vivek Nair using ``MetaGuard,'' our implementation of incognito mode for VR. (From `Going Incognito in the Metaverse,'$^{12}$ with permission.)}\vspace*{-5pt}
\end{figure*}

\section{SAFEGUARDING MOTION}
Data privacy issues are not unique to the metaverse. In fact, nearly every major communications technology advancement of the past century has been accompanied by corresponding privacy risks, from the wiretapping of landlines beginning in the 1890s through to emerging privacy concerns with smart home, mobile, and wearable devices today. As with XR motion, data that exists to provide necessary, legitimate functionality can often be leveraged for adversarial purposes.

On the web, tracking cookies are a quintessential example of this phenomenon. While cookies serve an important, legitimate purpose in enabling persistent sessions, adversaries can leverage them to track users across websites. But unlike in VR, the maturation of web technologies has brought a suite of countermeasures to such attacks. Technologies like VPNs, proxies, Tor, and incognito mode in browsers, have provided users with vital defensive tools for reclaiming their privacy in the face of such attacks. As of now, no equivalent comprehensive defensive utilities have been developed for extended reality devices.

We find ourselves now in the dangerous situation of facing unprecedented privacy threats in VR while lacking the defensive resources we have become accustomed to on the web. This is not necessarily due to a lack of interest in XR privacy, though research in this area is certainly far less common today than in web security and privacy. Rather, it is due to a fundamental challenge with XR motion data: the same telemetry data that is necessary to provide legitimate multi-user functionality can also be used for adversarial purposes.

Consider, by contrast, the permission-based model used by a typical smartphone application. The data sources accessible to each application are segmented into discrete permissions, which must be granted to the application by the end-user on an individual basis. If a navigation app requests access to view a user's GPS location, they might approve the request based on the understanding that the application needs this information to function. However, if it instead asks to see their contact list or listen to their microphone, this would reasonably raise a red flag in the context of what the application actually needs. Not so in VR, where there is fundamentally just a single stream of motion telemetry data that is used by all applications for a variety of purposes. Once sent off to a remote game server, there is no easy way to audit whether the data is being used for benign or nefarious reasons.

There is a silver lining, in that we have the opportunity to learn from the most effective privacy-preserving technologies on the web to implement metaverse architectures with security and privacy at their core. There are a few potential paths forward in this respect.

The first and most obvious approach would be to leverage ``local epsilon-differential privacy,'' a statistical measure of information leakage that is known as the ``gold standard of data privacy.'' We have already had moderate success in utilizing this technique. In a fourth study, we developed ``MetaGuard,'' an open-source plugin for the Unity game engine that we think of as a proof of concept for an ``incognito mode of the metaverse.'' MetaGuard works by identifying a number of privacy-sensitive dimensions present in an XR telemetry data stream, such as those corresponding to a user's height or wingspan. These axes are then passed through a ``Laplacian noise distribution,'' a type of differentially-private transformation function, before being transmitted to the server and on to other users. The plugin can easily be installed by end users into a variety of existing VR applications just by placing the extension files in a particular directory on their device, and can be customized to suit the specific needs and risks of each application, as shown in Figure 4.

To evaluate the efficacy of MetaGuard at protecting VR users, we replayed the motion recordings of users from the MetaData study, as well as of the 55,000 Beat Saber users, within a virtual environment to simulate what their data would have looked like had they been using MetaGuard. We found that MetaGuard is reasonably effective at mitigating both identification and inference attacks. MetaGuard reduced the accuracy of identifying users across sessions from nearly 95\% to less than 5\%. Attacks aiming to infer private user data were even further hindered, with the ability to infer demographics like age and gender dropping below the threshold of statistical significance.$^{12}$ 

\eject

These countermeasures do come at the cost of usability, however; by changing the user's motions to protect their privacy, users may experience a discrepancy between their true and apparent joint locations. For example, when reaching out to shake the hand of another virtual user, a person may find that the other user perceives their hand to be in a different location than expected, in order to hide their true wingspan from the other user. However, the level of ``error'' experienced by users is distributed according to a theoretical optimality that minimizes the error experienced by users for any given level of privacy.

Machine learning may provide an alternative approach to differential privacy for removing sensitive data from XR telemetry. A class of machine learning architectures are currently being developed to transform, or ``corrupt,'' data streams in order to remove user data embedded in the stream while minimally impacting legitimate application functionality. These architectures are typically trained using an adversarial network,$^{13}$ whereby one model attempts to infer private user data from a data stream, while another model tries to ``trick'' the first model without degrading the user experience. In using these types of models, we lose the mathematically provable properties of a differential privacy oriented approach, as formal verification on complex machine learning models remains a known difficult problem. On the flip side, the models could actually be more effective at protecting user privacy in practice, due to their ability to detect and obscure not only primary sensitive attributes but also hidden correlations to these variables. For example, if instructed to hide a user's height, the model would quickly learn to also obscure their wingspan, which is highly correlated to the former. As such, we consider this type of privacy-preserving approach to be an important next step in our research of metaverse security.

A third and final defense worth exploring is the use of trusted execution environments (TEEs) or secure multi-party computation (MPC) to provide transparency into how metaverse servers actually utilize the telemetry data shared by users. TEEs like Intel's SGX or Amazon's Nitro provide a hardware-based attestation mechanism that allows users to verify the software running on a remote machine before sending their data to that server, ensuring that only legitimate operations are being performed. For a subset of the operations offered by TEEs, MPC can also provide a purely cryptographic mechanism for achieving the same verifiable computations, regardless of the underlying hardware. These solutions are also not without their fair share of concerns. Most forms of MPC are currently far too inefficient to facilitate the high-throughput and low-latency data streams required for XR. TEEs, on the other hand, are fast enough, but researchers constantly demonstrate new security vulnerabilities that undermine their fundamental security properties. Still, technologies that enable users to audit exactly how their data is being used by metaverse entities may ultimately prove more resilient than motion transformation methods that cannot provide strong guarantees against an adaptive adversary that develops new ways to attack XR data streams over time.

\section{MOVING FORWARD}
XR technology is currently on track to become a ubiquitous means of accessing the internet, with AR devices having the potential to replace most of the existing portable electronic devices a consumer would typically carry today. With the forthcoming introduction of Apple into the XR device market, plus tens of billions of dollars in annual research and development expenditure from existing players like Meta, Microsoft, Google, Valve, and HTC, some of the largest and most influential technology companies on earth are clearly betting big on XR playing a significant role in the future of human-computer interaction.

Given that several of the major players in the metaverse space have their roots in advertising, the temptation will surely exist to leverage existing sales channels to monetize metaverse user data. Thus, we are currently at a crossroads. If nothing is done to improve the metaverse's present security and privacy posture, it is poised to inherit an exaggerated version of the privacy issues that are prevalent on the web. However, if we take the opportunity to learn from the history of browser-based attacks and defenses, security and privacy practitioners can prioritize research in this field and build privacy-preserving mechanisms into the fabric of the metaverse before the theoretical threats actually become widespread.

\vspace{-1em}
\section{ACKNOWLEDGMENTS}

We appreciate the support of Atticus Cull, Allen Yang, Beni Issler, Björn Hartmann, Brandon Huang, Charles Dove, Christopher Harth-Kitzerow, Eric Wallace, Gonzalo Munilla Garrido, Ines Bouissou, James Smith, Jason Sun, Julien Piet, Justus Mattern, Lun Wang, Mark Roman Miller, Rui Wang, Shuixian Li, Sriram Sridhar, Syomantak Chaudhuri, Wenbo Guo, Xiaoyuan Liu, and Yu Gai in the construction of this article and the original research supporting this work.
This work was supported in part by the National Science Foundation, the National Physical Science Consortium, the Fannie and John Hertz Foundation, and the Berkeley Center for Responsible, Decentralized Intelligence.
\vspace{-1em}

\def\refname{REFERENCES}

\vspace*{-8pt}

\eject

\begin{IEEEbiography}{Vivek Nair}{\,} is a Computer Science Ph.D. Student at the University of California, Berkeley, Berkeley, CA, USA. His current research interests include computer security and privacy, applied cryptography, and user authentication. Nair received the Master's degree in Computer Science from the University of Illinois at Urbana-Champaign. He is a Fellow of the National Science Foundation, the National Physical Science Consortium, and the Fannie and John Hertz Foundation. Contact him at vcn@berkeley.edu.\vspace*{8pt}
\end{IEEEbiography}

\begin{IEEEbiography}{Louis Rosenberg} {\,} is the CEO and Chief Scientist of Unanimous AI, Pismo Beach, CA, USA. His research interests include augmented reality, virtual reality, and artificial intelligence. Rosenberg received the Ph.D. degree in Mechanical Engineering from Stanford University. He is a Global Technology Advisor of the XR Safety Initiative and the Chief Scientist of the Responsible Metaverse Alliance. Contact him at louis@unanimous.ai.\vspace*{8pt}
\end{IEEEbiography}

\vfill
\eject

\begin{IEEEbiography}{James F. O'Brien}{\,} is a Professor of Computer Science at the University of California, Berkeley, Berkeley, CA, USA. His current research interests include computer graphics, animation, simulation, virtual reality, and machine learning. O'Brien received the Ph.D. degree in Computer Science from the Georgia Institute of Technology. He is a Fellow of the Alfred P. Sloan Foundation, the Okawa Foundation, and the Hellman Foundation. Contact him at job@berkeley.edu.\vspace*{8pt}
\end{IEEEbiography}

\begin{IEEEbiography}{Dawn Song} {\,} is a Professor of Computer Science at the University of California, Berkeley, Berkeley, CA, USA. Her current research interests include deep learning, computer security, and blockchains. Song received the Ph.D. degree in Computer Science from the University of California, Berkeley. She is a Fellow of the John D. and Catherine T. MacArthur Foundation, the Alfred P. Sloan Foundation, and the John Simon Guggenheim Memorial Foundation. Contact her at dawnsong@berkeley.edu.\vspace*{8pt}
\end{IEEEbiography}

\end{document}